\documentclass[pra,twocolumn,floatfix,reprint,footinbib,superscriptaddress,longbibliography]{revtex4-2}

\usepackage{amsmath,amssymb,amsthm,amstext,amsbsy,mathtools}
\usepackage{graphicx}
\usepackage{dcolumn}
\usepackage{mdwlist}
\usepackage{booktabs,multirow,threeparttable}
\usepackage{textcomp}
\usepackage[caption=false]{subfig}
\usepackage{appendix}
\usepackage{soul}
\usepackage{bbm}
\usepackage{svg}
\usepackage{mathrsfs,upgreek,braket}
\usepackage{xcolor}
\usepackage[colorlinks]{hyperref}
\usepackage[T1]{fontenc}
\usepackage{lmodern}
\usepackage{microtype}
\hypersetup{
    plainpages=true,
    pageanchor=true,
    colorlinks=true,
    breaklinks=true,
    hidelinks=false,
    linkcolor={blue},
    citecolor={red},
    urlcolor={blue},
    anchorcolor={black},
    hypertexnames=false
}
\definecolor{Dgreen}{RGB}{0,100,0}

\begin{document}
    
    \title{Enhancing light-matter coupling for exploring chaos in the quantum Rabi model}
    \author{Yan-Song Hu}
    \affiliation{Fujian Key Laboratory of Quantum Information and Quantum Optics, College of Physics and Information Engineering, Fuzhou University, Fuzhou 350108, China}

    \author{Yuan Qiu}
    \affiliation{Fujian Key Laboratory of Quantum Information and Quantum Optics, College of Physics and Information Engineering, Fuzhou University, Fuzhou 350108, China}
    
    \author{Ye-Hong Chen}\thanks{yehong.chen@fzu.edu.cn}
	\affiliation{Fujian Key Laboratory of Quantum Information and Quantum Optics, College of Physics and Information Engineering, Fuzhou University, Fuzhou 350108, China}
    \affiliation{Quantum Information Physics Theory Research Team, Center for Quantum Computing, RIKEN, Wako-shi, Saitama 351-0198, Japan}

    \author{XinYu Zhao}
    \affiliation{Fujian Key Laboratory of Quantum Information and Quantum Optics, College of Physics and Information Engineering, Fuzhou University, Fuzhou 350108, China}

    \author{Yan Xia}\thanks{xia-208@163.com}
	\affiliation{Fujian Key Laboratory of Quantum Information and Quantum Optics, College of Physics and Information Engineering, Fuzhou University, Fuzhou 350108, China}
    
    \date{\today}
    
    \begin{abstract}
        Accessing chaos in the quantum Rabi model (QRM) usually requires operating far from resonance, combined with ultra- or deep-strong light-matter coupling.
        This makes direct experiments challenging.
        In this manuscript, we propose a solution to this challenge by employing an anti-squeezing transformation to the bosonic field.
        Specifically, we demonstrate that this transformation maps a weakly coupled, two-photon driven Jaynes-Cummings model (JCM) to an effective deep-strong-coupling QRM in the squeezed-light frame.
        Using out-of-time-order correlator, Husimi distribution, and linear entanglement entropy, we numerically probe chaos in this coupling-enhanced platform and observe the similar chaotic phenomena as in the ideal QRM.
        We also find the coupling-enhanced model can drive the system deeper into the chaotic regime. 
        This establishes coupling-enhanced method as a practical approach to study QRM chaos without requiring intrinsic ultra-strong coupling.
    \end{abstract}

    \maketitle

    \section{\textbf{Introduction}}\label{s1}
        Studies of chaos in light-matter systems have largely focused on the Dicke model~\cite{Dicke1954PR,deAguiar1992AP,Hou2004PRA,Emary2003PRE,Li2024PRA,Ray2016PRE,PerezFernandez2011PRE,Lobez2016PRE}.
        By contrast, the quantum Rabi model (QRM)~\cite{Rabi1936PR,Rabi1937PR,Klimov2009BOOK,Braak2011PRL,Xie2017JPA,Chen2024PRL,Yan2024PRA,Chen2024cp} has seen less systematic investigations of chaos.
        This is primarily due to its single-atom nature and the absence of a direct classical counterpart for Rabi oscillations~\cite{Casati1995BOOK,Haake1991BOOK}, which hinders the application of traditional chaos analysis.
        Nevertheless, recent studies~\cite{Kirkova2022PRA,Wang2025CTP} further indicate that when the ratio of the atomic to cavity frequencies in the QRM is sufficiently large, the model admits an effective semiclassical, many-body description, enabling the use of semiclassical tools to probe chaotic behavior in the QRM.
        Although the QRM has been implemented across a range of experimental platforms~\cite{Thompson1992PRL,wallraff2004nature,Viennot2015science,koch2023NC,Diaz2019RMP}, observing chaos in this system remains a significant experimental challenge, as it requires both ultra-strong coupling and operation in the deep-detuning regime~\cite{Kockum2019NRP,Diaz2019RMP,Casanova2010PRL,Beaudoin2011PRA}.

        A promising path forward has emerged with the concept of frame transformations.
        Previous works~\cite{Ballester2012PRX,lv2015PRL,Qin2018PRL,Leroux2018PRL,Chen2019PRA,Chen2021PRL,burd2021np} has shown that when external driving is introduced, a system described by the Jaynes-Cummings model (JCM)~\cite{Jaynes1963ProcIEEE,Shore1993JMO,Abdalla1990PA,Niemczyk2010NP,Fink2008Nature,Reithmaier2004Nature} with weak couplings in the laboratory frame can be mapped to the QRM in a suitable frame.
        It is worth emphasizing that the JCM is itself the rotating-wave approximation of the more fundamental QRM, yet this mapping effectively reverses the approximation under specific conditions~\cite{Scully1997Book}.
        Building on this idea, an anti-squeezing technique has been proposed to enhance the interaction strength in otherwise weakly coupled cavity quantum electrodynamics (CQED) platforms.
        This approach enables access to QRM physics in the squeezed-light frame even though the laboratory-frame dynamics remain those of a weak-coupling JCM, providing a more feasible indirect route.
        It also allows convenient tuning of the effective frequencies to meet the requirements for observing chaos.
        Motivated by these advantages, the present work develops a theoretical study that bridges coupling-enhanced JCM with the exploration of chaos in the QRM.

        In this work, we map a weak-coupling JCM to an effective QRM in the squeezed-light frame via anti-squeezing transformation controlled by the squeezing parameter.
        This mapping yields an enhanced coupling strength and the effective cavity frequency in the squeezed-light frame.
        Through numerical simulations of chaos signatures in coupling-enhanced JCM, we find that the Loschmidt echo~\cite{Zhu2019PRA,Cucchietti2003PRL,Gorin2006reports} and state fidelity are highly sensitive to both the enhancement-induced error term and the initial state, making them unreliable for identifying chaos in the ideal Rabi model.
        Therefore, we focus on the following more reliable signatures.
        The out-of-time-order correlator $F(t)$~\cite{Garttner2017NP,Swingle2018NP} exhibits exponential growth of $1-F(t)$ at early time before fidelity decays sharply, 
        The linear entanglement entropy~\cite{Hou2004PRA,Lobez2016PRE} remain insensitive to the error term even at low fidelity. 
        Finally, the Husimi distribution~\cite{Takahashi1985PRL} also shows insensitivity to this error term under low-fidelity conditions.
        Parametric amplification effectively enhances the chaotic features by increasing both the effective coupling strength and the frequency ratio, driving the system deeper into the chaotic regime.
       
        The paper is organized as follows. 
        Sec.~\ref{s2} develops the theoretical framework for coupling enhancement via an anti-squeezing transformation and presents the effective Hamiltonian in the squeezed-light frame. 
        Sec.~\ref{s3} discusses the selection of system parameters and the preparation of initial states, and examines the impact of the error term on the simulation via fidelity. 
        Sec.~\ref{s4} systematically investigates chaos indicators suitable for this coupling-enhanced platform, including the out-of-time-order correlator, linear entanglement entropy, and the Husimi distribution. 
        Finally, Sec.~\ref{s5} summarizes the results and discusses prospects for experimental realization.

    \section{Coupling enhancement}\label{s2}
        For a weakly coupled and near-resonant CQED system (dissipation is neglected), the dynamics is described by JCM~\cite{Zueco2009PRA,Irish2007PRL}.
        We consider a two-photon driven JCM with Hamiltonian (hereafter $\hbar=1$)
            \begin{eqnarray}
                \hat{H}&=&\frac{\omega_{a}}{2}\hat{\sigma}_{z}+\omega_{c}\hat{a}^{\dagger}\hat{a}+g(\hat{a}^{\dagger}\hat{\sigma}_{-}+\hat{a}\hat{\sigma}_{+})\nonumber \\
                        &&-\frac{\lambda}{2}[e^{-i\omega_{p}t}\hat{a}^{\dagger2}+e^{i\omega_{p}t}\hat{a}^{2}],
                \label{eq8}
            \end{eqnarray}
        where $\hat{\sigma}_{z}$ is the Pauli operator;$\hat{\sigma}_{+}$ denotes the atomic raising operator; $\hat{\sigma}_{-}$ corresponds to the atomic lowering operator;
        $\hat{a}$ is the annihilation operator of cavity field and $\hat{a}^{\dagger}\hat{a}$ is the photon number operator;
        $\omega_{a}$ and $\omega_{c}$ are the atom and the cavity frequencies, respectively; and $g$ is the coupling strength; $\lambda$ is the parametric drive amplitude; and $\omega_{p}$ is the frequency of drive field.
        Choosing a rotating frame at frequency $\omega_{p}/2$, the transformation operator is defined as
            \begin{eqnarray}
                \hat{U}_{R}=\exp[i\frac{\omega_{p}}{2}(\hat{a}^{\dagger}\hat{a}+\frac{\hat{\sigma}_{z}}{2})t].
                \label{eq9}
            \end{eqnarray}
        In this frame, the Hamiltonian from Eq.~(\ref{eq8}) becomes
            \begin{eqnarray}
                \hat{H}^{R}&=&\frac{\delta_{a}}{2}\hat{\sigma}_{z}+\delta_{c}\hat{a}^{\dagger}\hat{a}+g(\hat{a}^{\dagger}\hat{\sigma}_{-}+\hat{a}\hat{\sigma}_{+})\nonumber \\
                                   &&-\frac{\lambda}{2}(\hat{a}^{\dagger2}+\hat{a}^{2}).
                \label{eq10}
            \end{eqnarray}
        where $\delta_{a(c)}=\omega_{a(c)}-\omega_{p}/2$.
        
        The rotated Hamiltonian is further transformed by applying the squeezing operator~\cite{Leroux2018PRL,Yu2025OE,Qiu2025NJP}
            \begin{eqnarray}
                \hat{U}_{S}[r]=\exp[r(\hat{a}^{2}-\hat{a}^{\dagger2})/2],
                \label{eq11}
            \end{eqnarray}
        with the squeezing parameter $r$ defined via $\tanh 2r=\lambda/\delta_{c}$.
        The effective Hamiltonian in the squeezed-light frame then becomes
            \begin{align}
                \hat{H}_{\text{eff}}=\hat{H}_{\text{Rabi}}+\hat{H}_{\text{err}},
                \label{eq12}
            \end{align}
        where
            \begin{eqnarray}
                \hat{H}_{\text{Rabi}}&=&\frac{\delta_{a}}{2}\hat{\sigma}_{z}+\Omega_{c}(r)\hat{a}^{\dagger}\hat{a}+\tilde{g}(\hat{a}^{\dagger}+\hat{a})(\hat{\sigma}_{+}+\hat{\sigma}_{-}),\nonumber \\
		        \hat{H}_{\text{err}}&=&-\frac{g}{2}e^{-r}(\hat{a}^{\dagger}-\hat{a})(\hat{\sigma}_{+}-\hat{\sigma}_{-}).%\nonumber \\
		        %\hat{H}_{\text{DA}} &=&-\frac{i\dot{r}(t)}{2}(\hat{a}^{\dagger2}-\hat{a}^{2}),
                \label{eq13}
            \end{eqnarray}
        Here, $\tilde{g}=ge^{r}/2$ is the enhanced coupling strength and $\Omega_{c}(r)=\delta_{c}\text{sech}2r$ is the effective cavity frequency.

    \section{Phase-space Mapping and Fidelity}\label{s3}           
        The choice of system parameters, initial states, and the effect of perturbations all significantly influence chaotic behavior.
        The system parameters determine whether chaos can be observed.
        Furthermore, studies on quantum-classical correspondence indicate that, within the classical approximation, initial quantum states associated with chaotic regions of phase space exhibit stronger sensitivity to perturbations and more distinct signatures of quantum chaos compared to those originating from regular regions~\cite{Wang2025CTP,Zhu2019PRA,Takahashi1985PRL}.
        It is also established that hypersensitivity to perturbation is a key indicator of chaos~\cite{Quan2006PRL,Zhu2019PRA}.
        Therefore, although the error Hamiltonian $\hat{H}_{\text{err}}$ (Eq.~\ref{eq13}) is exponentially suppressed by $e^{-r}$, it remains necessary to examine how this term, which acts as a perturbation to the ideal Rabi model, affects the fidelity when $r$ becomes sufficiently large.

        In this section, we first outline the required relations among the system parameters. We then select appropriate initial states for systems with different parameter configurations. Finally, we compute the Loschmidt echo for the ideal Rabi model and analyze the long-time fidelity as a function of $r$ to assess the impact of the error term.
        
        \subsection{System parameters and initial states}
        The conditions for observing chaos in the QRM are primarily governed by two key parameters, namely the ratio $\eta$ and the critical coupling strength $g_{c}$.
        Following Ref.~\cite{Kirkova2022PRA}, we define
            \begin{eqnarray}
                 \eta =\frac{\delta_{a}}{\Omega_{c}(r)}=\frac{\delta_{a}}{\delta_{c}}\text{cosh}2r 
            \label{eq27}
            \end{eqnarray}
        which compares the qubit detuning $\delta_{a}$ to the effective cavity frequency $\Omega_{c}(r)$.
        In the limit $\eta\rightarrow \infty$, the out-of-time-order correlator (OTOC) of QRM exhibits exponential growth of $1-F(t)$ in the superradiant phase~\cite{Swingle2018NP,Garttner2017NP}.
        A detailed discussion of the OTOC is provided in Sec.~\ref{s4}.
        The transition between the normal and superradiant phase occurs at the critical coupling
            \begin{align}
                g_{c}=\frac{\sqrt{\delta_{a}\Omega_{c}(r)}}{2}.
                \label{eq28}
            \end{align}
        Notablely, when $\eta>18$, the system can be accurately described within a semiclassical many-body framework~\cite{Wang2025CTP}, illustrating that the QRM, even with only a single qubit, can emulate collective quantum many-body behavior.
        Thus, we can use a semiclassical approach to select initial states through the classical Poincaré section.
        
        The initial state is chosen as the tensor product of a Bloch coherent state (a coherent superposition of a spin-1/2 system) and a Glauber coherent state,
            \begin{eqnarray}
                \ket{\psi(0)}=\ket{\tau}\otimes\ket{\beta},
                \label{eq15}
            \end{eqnarray}
        where $\ket{\tau}$ is the Bloch coherent state for the qubit
            \begin{eqnarray}
                \ket{\tau}=\frac{1}{\sqrt{1+|\tau|^{2}}}(\ket{G}+\tau\ket{E}),
                \label{eq16}
            \end{eqnarray}
        where $\ket{G}$ and $\ket{E}$ are the ground and excited eigen state of Pauli operator $\hat{\sigma}_{z}$, respectively.
        $\ket{\beta}$ is the Glauber coherent state for the cavity field in the squeezed-light frame,
            \begin{align}
                \ket{\beta}=e^{-|\beta|^{2}/2}\sum_{n = 0}^{\infty}  \frac{\beta^{n}}{\sqrt{n!}}\ket{n},
                \label{eq17}
            \end{align}
        where $\ket{n}$ is the $n$-photon Fock state in the squeezed-light frame.
        In the following, all cavity states are discussed within the squeezed-light frame.   
        This product state is chosen because it minimizes quantum uncertainty and most closely resembles a classical point in phase space, thereby providing a clear classical analog.
        The parameters of the coherent states map to phase-space coordinates via the following relations
            \begin{align}
                    q_1&=\sqrt{\frac{2}{1+\left\lvert \tau\right\rvert^2 }}\text{Re}(\tau),\nonumber\\
                    p_1&=\sqrt{\frac{2}{1+\left\lvert \tau\right\rvert^2 }}\text{Im}(\tau),\nonumber\\
                    q_2&=\sqrt{2}\text{Re}(\beta),\nonumber\\
                    p_2&=\sqrt{2}\text{Im}(\beta),
            \end{align}    
        where $(q_{1},p_{1})$ and $(q_{2},p_{2})$ are the phase-space coordinates of the qubit and field, respectively.
        
\begin{figure}[h]
            \centering
            \includegraphics[width=\columnwidth]{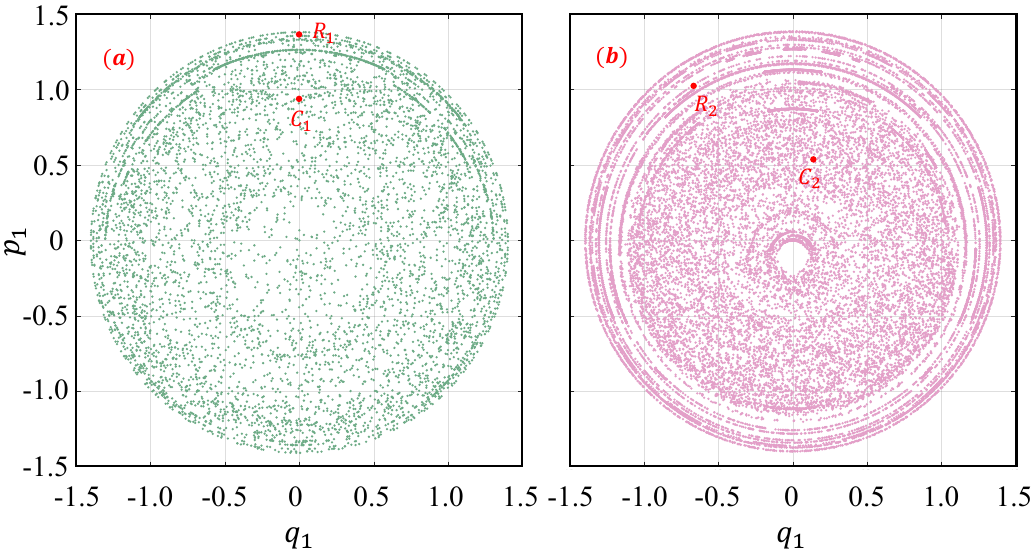}
            \caption{
                Poincaré section $(q_{2}=0, p_{2}>0)$ of semiclassical effective Hamiltonian. 
                (a) System parameters are set to $\delta_{a}=0.02\delta_{c}$, $g=2\times10^{-4}\delta_{c}$, $r=4$, with energy $E=0.018\delta_{c}$.
                Points $C_{1}(\tau=0.825,\beta=5.4461i)$ and $R_{1}(\tau=7,\beta=3.5384i)$ represent the chaotic and regular trajectories, respectively.
                (b) System parameters are $\delta_{a}=0.75\delta_{c}$, $g=0.0375\delta_{c}$, $r=2$, and energy $E=0.75\delta_{c}$.
                Points $C_{2}(\tau=0.0999+0.4081i,\beta=5.3065i)$ and $R_{1}(\tau=-0.9419+1.4653i,\beta=3.9644i)$ represent the chaotic and regular trajectories, respectively.
                }
            \label{fig1}
        \end{figure}

        We plot the Poincaré section in Fig.~\ref{fig1} with different sets of system parameters, both of which satisfy the conditions mentioned above, and mark the selected initial states on it.
        For a detailed discussion of the relationship between the selection of initial states and the Poincaré section, see Appendix~\ref{app:C}.

        \begin{figure}[b]
            \centering
            \includegraphics[width=\columnwidth]{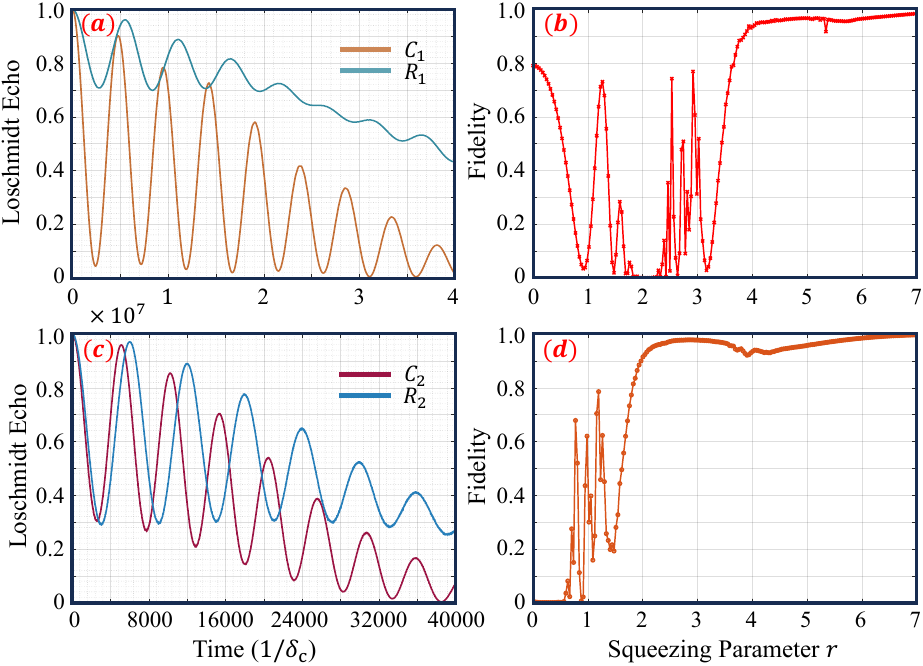}
            \caption{
                Time evolution of the Loschmidt echo and fidelity versus $r$.
                (a), (c) Time evolution of the Loschmidt echo for two different parameter sets.
                The system parameters and initial states are the same as in Figs.~\ref{fig1}(a) and Figs.~\ref{fig1}(b), respectively.
                (b), (d) Fidelity as a function of the squeezing parameter $r$ at a fixed time, with all other parameters and the initial states identical to those in (a) and (c), respectively.
                }
            \label{fig2}
            \end{figure}
           
        \begin{figure}[t]
            \centering
            \includegraphics[width=\columnwidth]{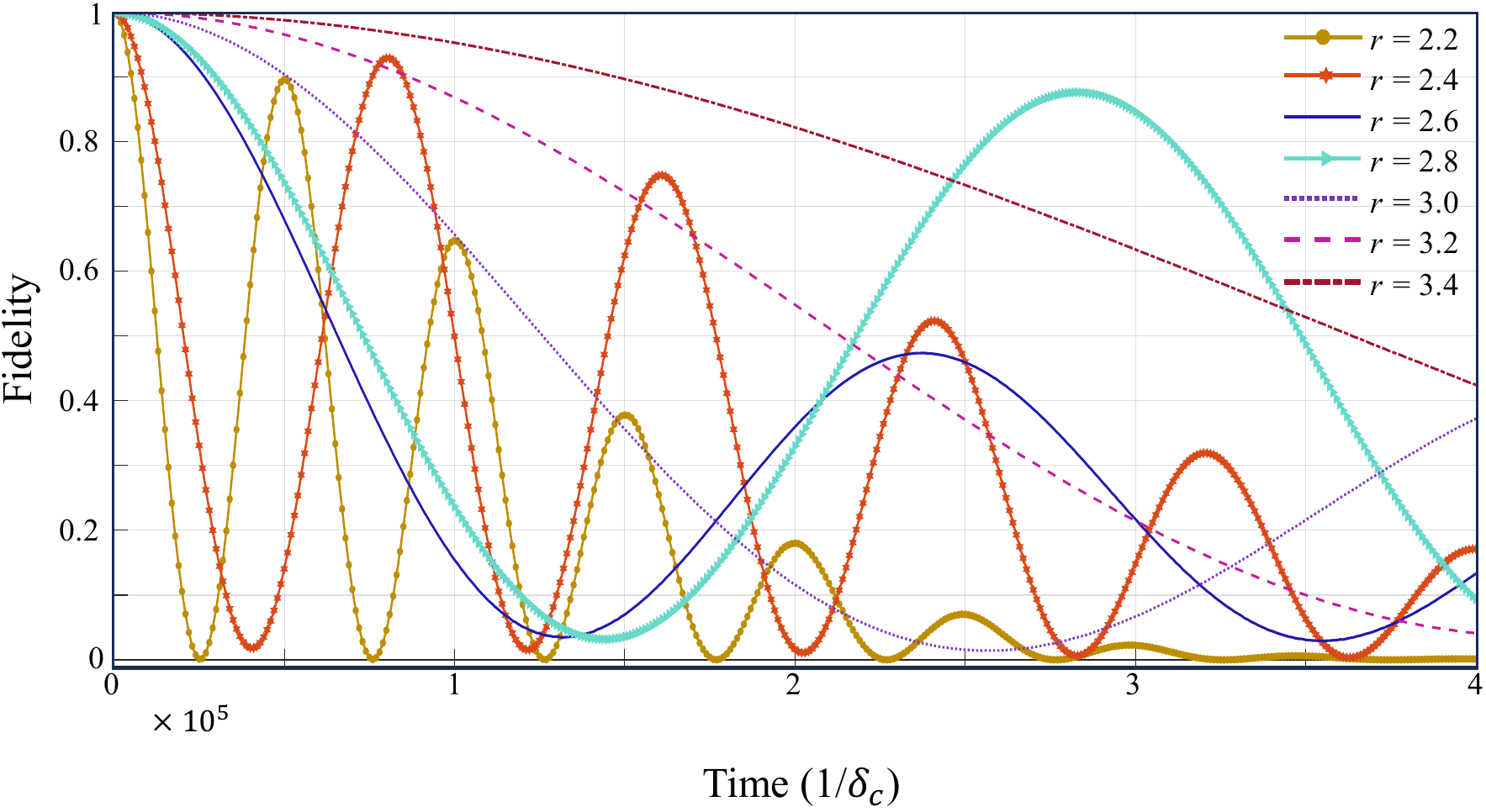}
            \caption{
                Time evolution of the fidelity for several representative squeezing parameters $r$.
                The system parameters and initial state are the same as in Fig.~\ref{fig2}(b).
                }
            \label{fig2.1}
            \end{figure}

        \subsection{Loschmidt Echo}
            We are interested in the conditions under which the dynamic of the Rabi model can be effectively simulated using coupling-enhanced CQED.
            To this end, we examine the fidelity between the evolution driven by the effective Hamiltonian $\hat{H}_{\text{eff}}$ and that driven by the ideal Rabi Hamiltonian $\hat{H}_{\text{Rabi}}$.
            The fidelity is defined as the overlap $|\langle \psi_{\text{eff}}(t)|\psi_{\text{Rabi}}(t)\rangle|^2 $ between $\ket{\psi_{\text{eff}}(t)}$ and $\ket{\psi_{\text{Rabi}}(t)}$.
            By choosing a suitable $r$, we can treat the error term $\hat{H}_{\text{err}}$ as a perturbation to the ideal Rabi Hamiltonian $\hat{H}_{\text{Rabi}}$, then the fidelity effectively corresponds to the Loschmidt echo of the ideal Rabi model.
            The Loschmidt echo quantifies the sensitivity of a quantum system to perturbations~\cite{Zhu2019PRA,Cucchietti2003PRL,Quan2006PRL} and is defined as
            \begin{align}
                L(t)=|\bra{\psi(0)}e^{i\hat{H}t}e^{-i(\hat{H}+\hat{H}_{\text{per}})t}\ket{\psi(0)}|^2, 
            \end{align}
            where $\hat{H}=\hat{H}_{\text{Rabi}}$ and $\hat{H}_{\text{per}}=\hat{H}_{\text{err}}$.
            
            In Fig.~\ref{fig2}(a) and Fig.~\ref{fig2}(c), we present the time evolution of the Loschmidt echo under different parameter sets, using the same system parameters and initial states as in Fig.~\ref{fig1}(a) and Fig.~\ref{fig1}(b), respectively.
            It can be observed that the initial state located in the chaotic region exhibits a faster fidelity decay, with its long-time fidelity evolution showing strongly damped oscillations.
            In panel (c), although $C_2$ and $R_2$ correspond to the chaotic and regular regions on the Poincaré section respectively, their Loschmidt echoes fail to exhibit a clear distinction.
            This indicates that the Loschmidt echo is also highly sensitive to the system parameters.

            In Fig.~\ref{fig2}(b) and Fig.~\ref{fig2}(d), we also plot the long-time fidelity as a function of the squeezing parameter $r$, while keeping all other system parameters fixed and identical to those in (a) and (c).
            In panel (b), the initial state is set to $C_1$ with the evolution time fixed at $T=4\times10^5\delta_{c}^{-1}$;in panel (d), the initial state is set to $C_2$ with $T=600\delta_{c}^{-1}$.
            As $r$ increases, the fidelity oscillate strongly.
            Once $r$ becomes sufficiently large, the fidelity increas sharply and stabilizes at a high value.
            However, owing to the high sensitivity of the Loschmidt echo to system parameters and initial states, a slight decrease in fidelity can even be observed with further increase of $r$.
            In Fig.~\ref{fig2.1}, we plot the time evolution of the fidelity for several values of $r$ corresponding to those in Fig.~\ref{fig2}(b).
            It can be seen that the erratic oscillations observed in figs.~\ref{fig2}(b) and (d) can be attributed to phase differences associated with different $r$ values.
            At a fixed time, these phase differences cause the fidelity to deviate from a monotonic increase with $r$, resulting instead in the sawtooth pattern displayed in Figs.~\ref{fig2}(b) and (d).

            Based on the results presented above, we have seen that the Loschmidt echos is highly sensitive to both the system and the choice of initial state.
            Furthermore, our simulations reveal that Loschmidt echos' response is also sensitive to different forms of perturbations.
            For certain parameter sets, it fails to clearly distinguish chaotic from regular points.
            Whether the Loschmidt echo can be reliably employed as a robust experimental signature of quantum chaos in QRM requires further investigation.
            Moreover, a larger $r$ ensures a higher fidelity over a longer duration, albeit it needs longer time to observe signatures of chaos.
            The time cost will be further discussed in Sec.~\ref{s4}.
            These imply that the choice of the squeezing parameter $r$ must be considered in the context of the system parameters and larger values are not always preferable.
            Therefore, in this work we adopt two complementary approaches to probe chaotic behavior: first, employing indicators that are weakly affected by perturbations, and second, using signatures of chaos that become visible before the fidelity decays significantly. 
   
    \section{Suitable Chaos Indicators for Coupling-Enhanced CQED}\label{s4}
        In this section, we numerically explore several signatures suitable for detecting quantum chaos in the coupling-enhanced CQED platform. 
        We first employ the OTOC as an indicator of quantum chaos.
        Our simulations show that the OTOC can reveal clear signatures of chaos over short time scales, making it particularly useful for systems with rapid fidelity decay or limited cavity lifetimes.
        We then examine the linear entanglement entropy, finding it is robust against the errorterms introduced by coupling enhancement and faithfully captures the chaotic dynamics of the ideal Rabi model.
        Finally, we analyze the phase-space dynamics using the Husimi distribution.
        This representation distinguishes chaotic initial states rapidly, and its long-time distributions remain largely insensitive to the error term.
       
        \subsection{Out-of-time-order correlator}
            \begin{figure}[b]
                \centering
                \includegraphics[width=\columnwidth]{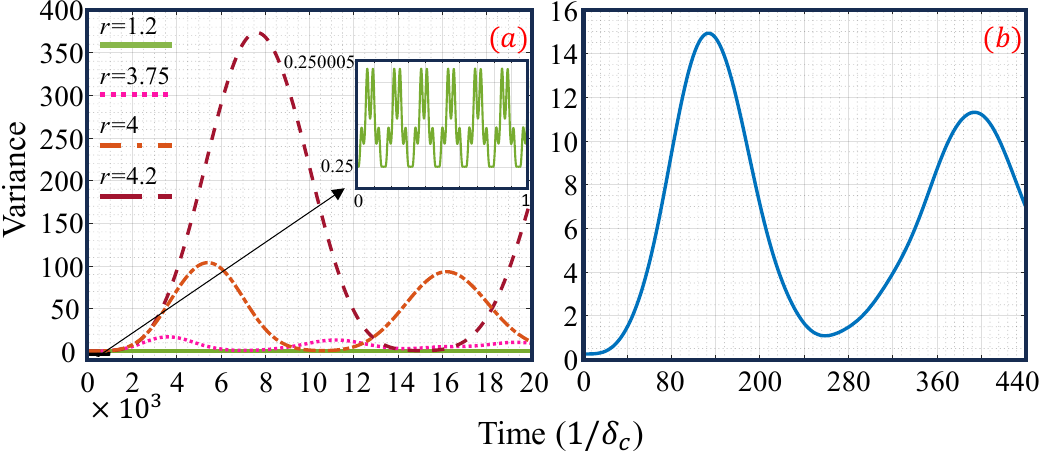}
                \caption{
                    Time evolution of $\text{var}[\hat{G}(t)]$ under different system parameters, with the initial state prepared as $\ket{\phi}=\ket{+}\otimes\ket{0}$.
                    (a) Dependence on the squeezing parameter $r$, while all other parameters are identical to those in Fig.~\ref{fig1}(a).
                    (b) System parameters are the same as in Fig.~\ref{fig1}(b).
                        }
                \label{fig3}
            \end{figure}
        The OTOC is defined as:
            \begin{align}
                F(t)=\bra{\phi}\hat{W}^{\dagger}(t)\hat{V}^{\dagger}\hat{W}(t)\hat{V}\ket{\phi},
                \label{eq25} 
            \end{align}
        where $\ket{\phi}$ is the initial state, $\hat{W}$ and $\hat{V}$ are two commuting operators $[\hat{W},\hat{V}]=0$ when $t=0$.
        Time-evolution is governed by the effective Hamiltonian as $\hat{W}(t)=e^{i\hat{H}_{\text{eff}}t}\hat{W}e^{-i\hat{H}_{\text{eff}}t}$.
        The OTOC quantify the scrambling of quantum information among different degrees of freedom in the system and serves as a standard diagnostic of quantum chaos.
        By invoking the correspondence between phase-space Poisson brackets and quantum commutators, we relate the OTOC with the classical Lyapunov exponent
            \begin{align}
                1-\text{Re}[F(t)]=\frac{1}{2}\langle [\hat{V}^{\dagger},\hat{W}^{\dagger}(t)][\hat{W}(t),\hat{V}]\rangle \sim e^{\lambda_{Q}t},
                \label{eq26}
            \end{align}
        where $\lambda_{Q}$ is quantum Lyapunov exponent, and has been conjectured to be a quantum analogue of a classical Lyapunov exponent~\cite{,Swingle2018NP,Kirkova2022PRA,xiang2025PRA}. 
        Here we choose the initial state $\ket{\phi}=\ket{+}\otimes\ket{0}~(\tau=1,\beta=0)$ which is the eigenstate of $\hat{\sigma}_{x}$.
        It has been demonstrated in Ref.~\cite{Kirkova2022PRA} that choosing $\hat{V}=\hat{\rho}(0)$ or $\hat{V}=\hat{\sigma}_{x}$ leads to similar growth patterns at early times in the system's evolution of QRM, so we select the same choice as those in Ref.~\cite{Kirkova2022PRA}.
        Then we take $\hat{V}=\hat{\rho}(0)=\ket{\phi}\bra{\phi}$ and $\hat{W}=e^{i\epsilon\hat{G}}$, where $\hat{G}=(\hat{a}+\hat{a}^{\dagger})/2$ and $\epsilon$ is a small perturbation.
        Expanding the OTOC in a power series of $\epsilon$, we obtain
            \begin{align}
                1-F(t)=\epsilon^2(\langle \hat{G}^2(t)\rangle-\langle \hat{G}(t) \rangle^2)=\epsilon^2\text{var}[\hat{G}(t)].
            \end{align}
        
        In Fig.~\ref{fig3}(a), we show how the time evolution of $\text{var}[\hat{G}(t)]$ depends on the squeezing parameter $r$. 
        The system parameters are fixed at $(\delta_{a}=0.02\delta_{c}$, $g=2\times10^{-4}\delta_{c})$ with the initial state $\ket{\phi}=\ket{+}\otimes\ket{0}$.
        During the early stage of the evolution, $\text{var}[\hat{G}(t)]$ is observed to grow exponentially for $r$ exceed $3.75$, reaching a maximum at the scrambling time $t^{*}$, which is defined as the time when $\text{var}[\hat{G}(t)]$ first attains its maximum value.
        From $t^{*}$, the lyapunov exponent can be extracted.
        In contrast, for $r=1.2$, $\text{var}[\hat{G}(t)]$ rapidly saturates at a much lower maximum value and subsequently exhibits fast periodic oscillations.
        These choices of $r$ make the systems correspond to the superradiant phase and the normal phase, respectively.
        Ref.~\cite{Kirkova2022PRA} shows that, in the superradiant phase of the ideal Rabi model with fixed critical coupling $g_{c}$, either increasing $\eta$ at fixed coupling strength or enhancing the coupling at fixed $\eta$ leads to a more pronounced exponential growth and a larger maximum value.
        Our coupling-enhancement scheme simultaneously increases both $\eta$ and the effective coupling strength, while lowering the threshold $g_{c}$ for entering the superradiant phase (see Eq.~\ref{eq27} and Eq.~\ref{eq28}), making the onset of chaos significantly easier to observe. 

        However, it can also be seen that simply increasing $r$ prolongs the scrambling time $t^{*}$, which implies that a longer evolution time is required to observe clear signatures of chaos.
        This places more stringent demands on the cavity lifetime.
        Moreover, since $\tanh2r=\lambda/\delta_{c}$, increasing $r$ drives the system closer to the instability threshold $\lambda=|\delta_{c}|$~\cite{Leroux2018PRL}, imposing tighter precision requirements on the experimental implementation.
        
        In Fig.~\ref{fig3}(b), We consider another parameter set ($\delta_{a}=0.75\delta_{c}$, $g=0.0375\delta_{c}$, $r=2$).
        An exponential growth emerges within a significantly shorter time window.
        Although Fig.~\ref{fig2}(c) shows that the corresponding fidelity remains high only for a limited duration, this time scale is already sufficient to observe the OTOC as a chaos indicator in the coupling-enhanced CQED platform.
        
            \begin{figure}[h]
                \centering
                \includegraphics[width=\columnwidth]{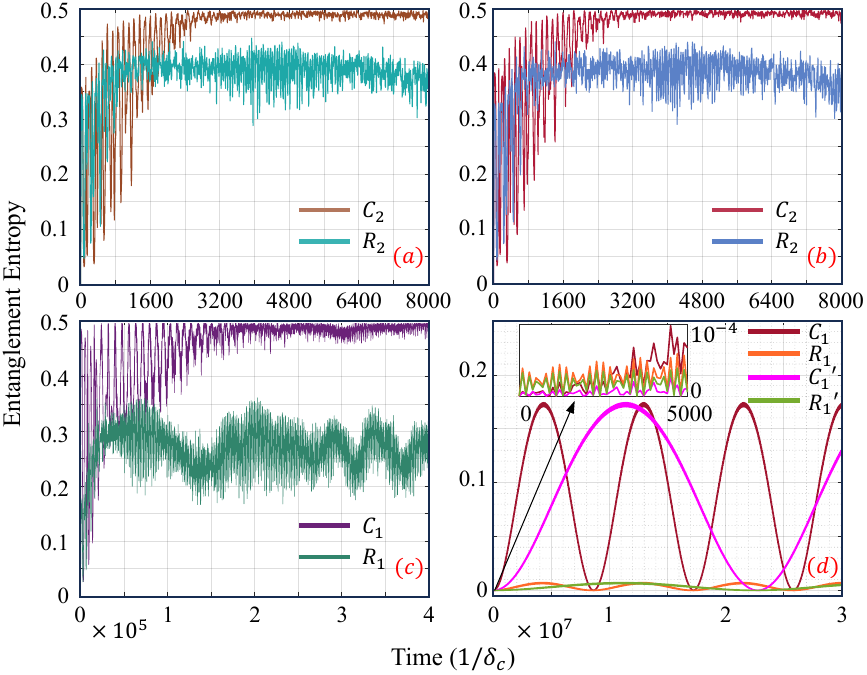}
                \caption{
                    Time evolution of the linear entanglement entropy.
                    (a), (b) System parameters and initial states are identical to those in Fig.~\ref{fig1}(b), with evolution driven by the effective Hamiltonian $\hat{H}_{\text{eff}}$ and the ideal Rabi Hamiltonian $\hat{H}_{\text{Rabi}}$, respectively.
                    (c) System parameters and initial states are the same as in Fig.~\ref{fig1}(a), under evolution driven by the effective Hamiltonian $\hat{H}_{\text{eff}}$.
                    (d) Squeezing parameter $r=1.2$; all other parameters match those in panel (c).
                    Evolutions for points $C_{1}^{'}$ and $R_{1}^{'}$ are driven by the ideal Rabi Hamiltonian $\hat{H}_{\text{Rabi}}$.
                    The inset shows an enlarged view of the entropy oscillations over a selected time interval.
                    }
                \label{fig4}
            \end{figure}
        \subsection{Linear Entanglement Entropy}
        Linear entanglement entropy is used to describe the degree of entanglement between subsystems and the entire system.
        Supposing that we have a full system consists of two subsystems, if we want to focus on the subsystem 1, the linear entanglement entropy is defined as
            \begin{eqnarray}
                S(t)=1-\text{Tr}[\hat{\rho}_{1}(t)^2],
                \label{eq23}
            \end{eqnarray}
        where $\hat{\rho}_{1}$ is the reduced density matrix of the subsystem 1 (the atomic part), obtained by tracing out the cavity subsystem from the full density matrix $\hat{\rho}=\ket{\psi(t)}\bra{\psi(t)}$
            \begin{eqnarray}
                \hat{\rho}_{1}=\text{Tr}_{2}(\hat{\rho}).
                \label{eq22}
            \end{eqnarray}
        For a pure, separable total state, the reduced density matrix is also pure, satisfying $\mathrm{Tr}(\hat{\rho}^{2})=1$ and $S=0$; by contrast entanglement renders $\hat{\rho}_{1}$ mixed, yeilding $\mathrm{Tr}(\hat{\rho}^{2})<1$ and $S>0$.
        Larger $S$ indicates a higher degree of mixing in the subsystem and strong entanglement between the two subsystems.
        Entanglement has been proposed and validated as an indicator of quantum chaos~\cite{Wang2004PRE}, and linear entropy has been successfully applied in the Dicke model~\cite{Hou2004PRA,Lobez2016PRE}.
        Here we use it to probe chaos in coupling-enhanced CQED with different system parameters and initial states, numerically.

        In Fig.~\ref{fig4}(a) and Fig.~\ref{fig4}(b), we plot the time evolution of the linear entanglement entropy driven by the ideal Rabi Hamiltonian $\hat{H}_{\text{Rabi}}$ and the effective Hamiltonian $\hat{H}_{\text{eff}}$, respectively, to examine the influence of the error term $\hat{H}_{\text{err}}$ on the entanglement entropy.
        The system parameters and the choice of initial state are the same as those in Fig.~\ref{fig1}(b).
        It can be observed that the fidelity remains at a low level for most of the evolution time (see Fig.~\ref{fig2}(c)).
        However, this does not affect the overall trend of the entanglement entropy evolution, influencing only finer-scale oscillations.
        The saturation value of the entanglement entropy for the chaotic initial point $C_2$ is markedly higher than that for the regular point $R_2$.
        Moreover, the entanglement entropy at the chaotic point stabilizes at high level, exhibiting only smaller-amplitude oscillations.
        In Fig.~\ref{fig4}(c), using the parameters and initial state form Fig.~\ref{fig1}(a), we plot only the entanglement entropy evolution under the effective Hamiltonian, as the fidelity remains high throughout the evolution.
        A more pronounced distinction between $C_1$ and $R_1$ can be observed in this case.
        
\begin{figure}
            \centering
                \includegraphics[width=\columnwidth]{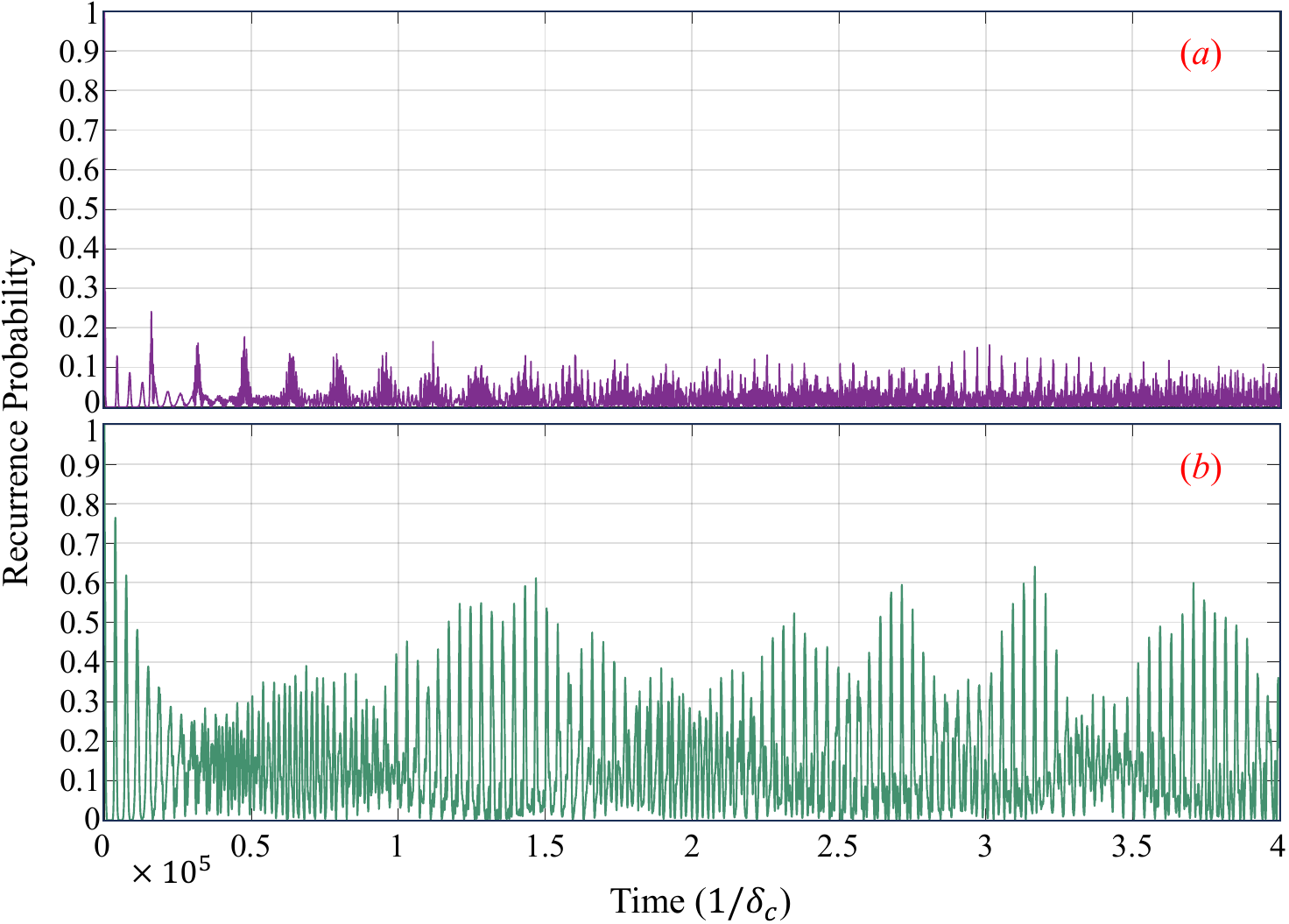}
                \caption{
                    Time evolution of recurrence probability.
                    System parameters are identical to those in Fig.~\ref{fig4}(c), with evolution driven by the effective Hamiltonian $\hat{H}_{\text{eff}}$.
                    (a) The initial state is chosen as $C_1$.
                    (b) The initial state is chosen as $R_1$.
                    }
                \label{fig4.1}
        \end{figure}

        The rapid oscillations of the entanglement entropy shown in the figure can be explained by the recurrence probability, which illustrates the quantum collapse and partial revival~\cite{Casanova2010PRL,Naqvi2000PS,Peres1993PRA} and is defined as
        \begin{align}
            P(t)=|\bra{\psi(0)}\ket{\psi(t)}|^2.
        \end{align}
        We then select two initial states from Fig.~\ref{fig4}(c) and plot the time evolution of their recurrence probabilities in Fig.~\ref{fig4.1}.
        It can be seen that the initial state $R_1$ exhibits partial revivals with larger amplitude during the evolution, while $C_1$ shows smaller amplitude.
        The recurrence probabilities of these two initial states display the same difference in amplitude as their entanglement entropies.
        Moreover, the frequency of the collapses and partial revivals matches the oscillation frequency of the entanglement entropy, indicating that the oscillations in entanglement entropy arise from changes in the degree of entanglement caused by collapses and partial revivals.
        Incidentally, quantum collapses and revival have also been employed in previous studies to investigate chaotic behavior.

        Then we consider the parameter set corresponding to the normal phase (see Fig.~\ref{fig3}(a)), fixing $r=1.2$ while keeping all other parameters and the initial states the same as in panel (c).
        Under both the effective Hamiltonian and the ideal Rabi Hamiltonian, different initial states exhibit similarly rapid, small-amplitude oscillations superimposed on a slow envelope.
        In this regime, the error term cannot be neglected.
        However, it does not affect the maximum value of the entanglement entropy but only modifies its oscillation frequency.
        We plot the time-averaged entanglement entropy to characterize the chaotic behavior of different initial states in Fig.~\ref{fig1}.
        
\begin{figure}
                \centering      
                \includegraphics[width=\columnwidth]{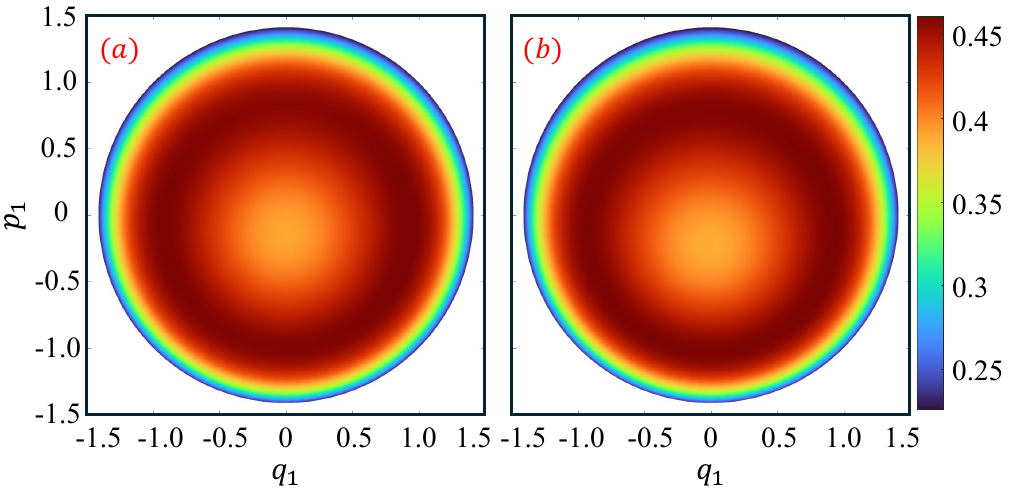} 
                \caption{
                    Distribution of the average entanglement on the Poincaré section.
                    (a) System parameters are identical to those in Fig.~\ref{fig1}(a), with an evolution time $T=4\times10^{5}\delta_{c}^{-1}$.
                    (b) System parameters are identical to those in Fig.~\ref{fig1}(b), with an evolution time $T=8000\delta_{c}^{-1}$.
                        }
                \label{fig5}
            \end{figure}

        Averaging over time,
            \begin{eqnarray}
                \bar{S}=\frac{1}{T}\int_{0}^{T}S(t)dt,
                \label{eq24}
            \end{eqnarray}
        we find that stronger chaos correlates with larger $\bar{S}$ (Fig.~\ref{fig5}).
        The phase-space distribution of average entanglement entropy on the atomic subspace mirrors the chaotic sea and stable islands seen in the classical Poincaré sections (Fig.~\ref{fig1}): stable islands correspond to regions of lower $\bar{S}$, while a continuous variation of $\bar{S}$ appears across the  transition between regular and chaotic domains-analogous to sampling points becoming confined to progressively narrower rings before settling on invariant curves.
        
        We conclude that the linear entanglement entropy is robust against the error term $\hat{H}_{\text{err}}$ and can be readily measured in experiments, making it a reliable indicator across different parameter sets.
        Its main limitations are twofold: it cannot clearly distinguish initial states near the edge between chaotic and regular regions~\cite{Ray2016PRE,Wang2004PRE}, and it requires a longer cavity lifetime to be fully resolved in evolution.
       
        \begin{figure*}[t]
                \centering
                \includegraphics[width=\textwidth]{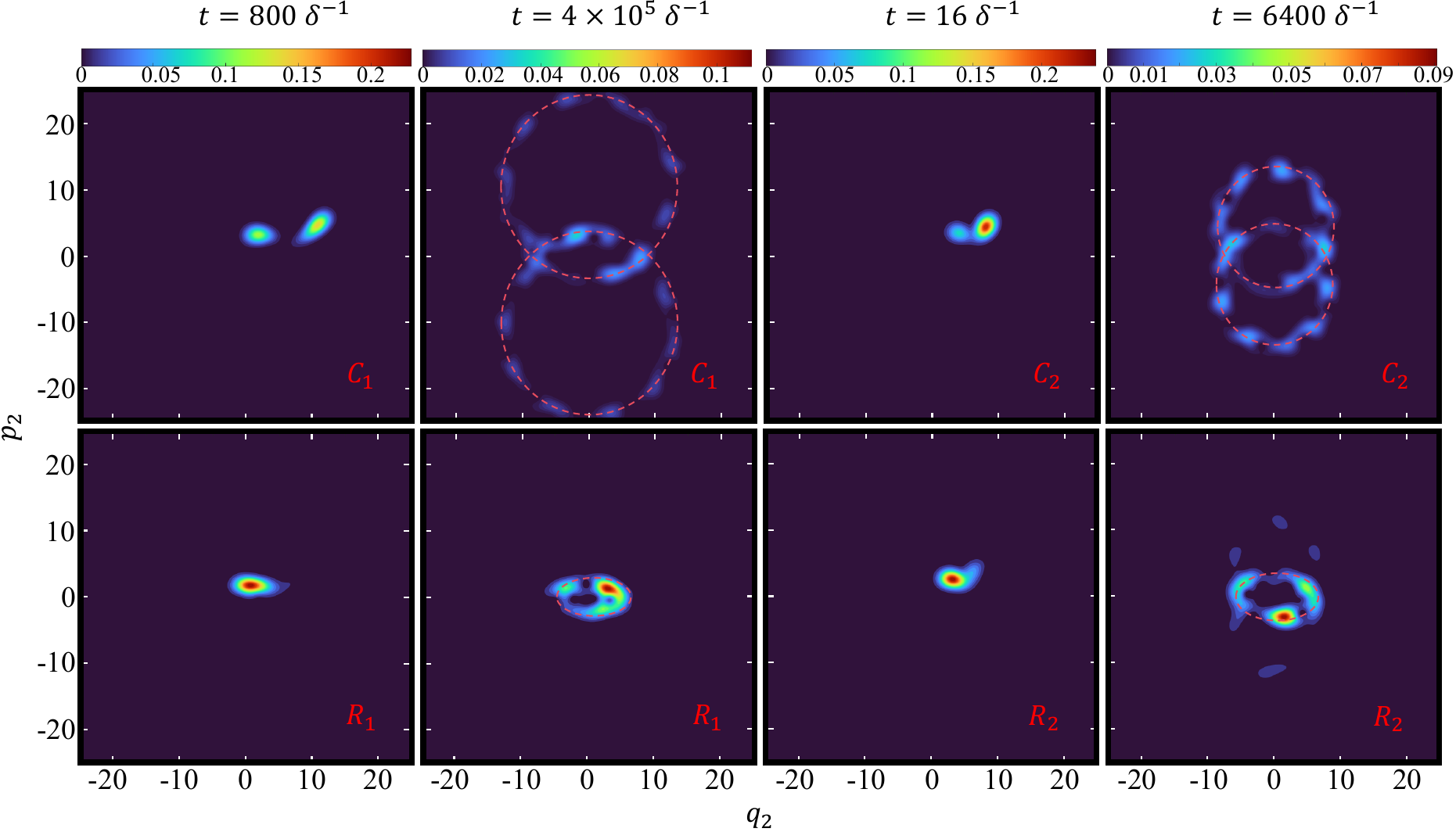}
                \caption{
                        Husimi distributions at different evolution times for two parameter sets.
                        Each column corresponds to a specific time snapshot.
                        The first two columns share the same system parameters and initial state as Fig.~\ref{fig1}(a), while the last two columns correspond to those in Fig.~\ref{fig1}(b).
                        }
                \label{fig6}
            \end{figure*}

        \subsection{Husimi Distribution}
        Visualizing the evolution of quantum states in phase space provides an important approach to studying quantum chaos.
        Altohugh the Wigner distribution is widely used for this purpose, it can take negative values and therefore does not constitute a genuine probability distribution.
        A physically motivated alternative is obtained by applying Gaussian smoothing to the Wigner distribution, which yields the Husimi quasi-probability distribution function~\cite{Takahashi1985PRL}.
        Husimi distribution function can be expressed in a compact form as the projection of the quantum density matrix onto Glauber coherent states~\cite{Wang2025CTP}.
            \begin{eqnarray}
                Q=\frac{1}{\pi}\bra{\beta}\hat{\rho}_{2}\ket{\beta},
                \label{eq21}
            \end{eqnarray}
        where $\hat{\rho}_{2}$ is the reduced density matrix of the cavity field, obtained by tracing out the atomic subsystem from the full density matrix $\hat{\rho}$.
        This distribution gives an intuitive picture of how a quantum state localizes or spreads over the underlying classical phase-space structures.
        The Husimi distribution of initial coherent state appear as Gaussian wave packets centered at the phase-space coordinates $(q_2,p_2)$.
        
        In Fig.~\ref{fig6} we plot the Husimi distribution at selected times.
        It is observed that the wave packets corresponding to initial states in the chaotic regions ($C_1$ and $C_2$) split rapidly within a short time, whereas those in the regular regions require a considerably longer duration to exhibit noticeable spreading.
        After prolonged evolution, the Husimi distribution for chaotic points clearly develops a `double-ring' structure, while for regular points it remains confined to a smaller region, displaying a `single-ring' profile.
        An exception is $R_2$, whose distribution also extends toward the periphery; this can be attributed to its location in the transitional zone between regular and chaotic regions.
        Comparison of the results from these two parameter sets shows that the further away the system is from the threshold conditions for chaos [Eqs. (7) and (8)], the more distinct the dynamical behaviors of regular and chaotic points become.

        Furthermore, the inclusion of the error term $\hat{H}_{\text{err}}$ does not alter these characteristic ring-shaped patterns.
        This demonstrates that the Husimi distribution remains a valid and effective tool in coupling-enhanced CQED platform.

    \section{conclusions}\label{s5}
        We have demonstrated that a weakly coupled, two-photon driven Jaynes-Cummings model can be mapped onto an effective quantum Rabi model with exponentially enhanced light-matter coupling via an anti-squeezing transformation in the squeezed-light frame. 
        This coupling-enhanced platform enables the exploration of chaos in the quantum Rabi model without requiring intrinsic ultra- or deep-strong coupling. 
        Through comprehensive numerical simulations, we identify the out-of-time-order correlator, linear entanglement entropy, and Husimi distribution as reliable indicators of quantum chaos in this setting, even in the presence of the error term arising from the transformation. 
        Specifically, the out-of-time-order correlator can reveal chaotic behavior before the fidelity undergoes rapid decay, while the linear entanglement entropy and Husimi distribution exhibit robustness against the error term and are suitable for long-time chaos detection. 
        In contrast, the Loschmidt echo displays high sensitivity to system parameters and initial states, failing to clearly distinguish chaotic from regular regions under certain conditions, suggesting that its reliability as a chaos diagnostic requires further careful assessment. 
        Furthermore, we show that increasing the squeezing parameter $r$ not only simultaneously enhances the effective coupling strength and frequency ratio but also drives the system deeper into the chaotic regime. 
        However, this comes at the cost of prolonged scrambling times and imposes stricter demands on experimental precision and cavity lifetime. 
        Our results provide concrete parameter windows and diagnostic tools for the experimental realization of previously inaccessible chaos in the Rabi model within weakly coupled cavity quantum electrodynamics platforms, thereby establishing a practical bridge between theoretical models and experimental implementation.
      
    \begin{acknowledgments}
        Y.-H.C. was supported by the National Natural Science Foundation of China under Grant No. 12304390 and 12574386, the Fujian 100 Talents Program, and the Fujian Minjiang Scholar Program.
        Y. X. was supported by the National Natural Science Foundation of China under Grant No. 62471143, the Key Program of National Natural Science Foundation of Fujian Province under Grant No. 2024J02008, and the project from Fuzhou University under Grant No. JG2020001-2.
    \end{acknowledgments}
    
    \appendix
        \section{Enhancing coupling via anti-squeezing}\label{app:B}
            Let $\hat{U}_{R}$ be the rotating-frame unitary defined in Eq.~(\ref{eq9}).
            In this frame, the Hamiltonian $\hat{H}$ (Eq.~(\ref{eq8})) transforms as
                \begin{eqnarray}
                    \hat{H}'=\hat{U}_{R}(t)\hat{H}\hat{U}_{R}^{\dagger}(t)-i\hat{U}_{R}(t)\frac{d}{dt}\hat{U}_{R}^{\dagger}(t),
                    \label{b1}
                \end{eqnarray}
            with the compensating term
                \begin{eqnarray}
                    -i\hat{U}_{R}(t)\frac{d}{dt}\hat{U}_{R}^{\dagger}(t)=-\frac{\omega_{p}}{2}(\hat{a}^{\dagger}\hat{}+\frac{\hat{\sigma}_{z}}{2}).
                    \label{b2}
                \end{eqnarray}
            The corresponding operator transformations are
                \begin{align}
                    \hat{a} &\rightarrow e^{-i\frac{\omega_{p}}{2}t}\hat{a},\nonumber \\
                    \hat{a}^{\dagger} &\rightarrow e^{i\frac{\omega_{p}}{2}t}\hat{a}_{\dagger},\nonumber \\
                    \hat{\sigma}_{\pm} &\rightarrow e^{\pm i\frac{\omega_{p}}{2}t}\hat{\sigma}_{\pm},
                    \label{b3}
                \end{align}
            under which the qubit, cavity, and parametric-drive terms become
                \begin{equation}
                \begin{aligned}
                    \frac{\omega_{a}}{2}\hat{\sigma}_z \;&\to\;
                    \tfrac{1}{2}\!\left(\omega_{a}-\tfrac{\omega_{p}}{2}\right)\hat{\sigma}_z,\\
                    \omega_{c}\hat{a}^\dagger \hat{a} \;&\to\;
                    \left(\omega_{c}-\tfrac{\omega_{p}}{2}\right)\hat{a}^\dagger \hat{a},\\
                    -\tfrac{\lambda(t)}{2}\!\left[e^{-i\omega_{p} t}\hat{a}^{\dagger 2}
                    + e^{i\omega_{p} t}\hat{a}^{2}\right]
                    \;&\to\; -\tfrac{\lambda(t)}{2}\!\left(\hat{a}^{\dagger 2}+\hat{a}^{2}\right).
                    \end{aligned}
                    \label{b4}
                \end{equation}
            Therefore,
                \begin{eqnarray}
                    \hat{H}'&=&\frac{1}{2}(\omega_{a}-\frac{\omega_{p}}{2})\hat{\sigma}_{z}+(\omega_{c}-\frac{\omega_{p}}{2})\hat{a}^{\dagger}\hat{a}\nonumber \\
                              &&+g(\hat{a}^{\dagger}\hat{\sigma}_{-}+\hat{a}\hat{\sigma}_{+})-\frac{\lambda(t)}{2}[\hat{a}^{\dagger2}+\hat{a}^{2}].
                    \label{b5}
                \end{eqnarray}
            and, defining the detunings $\delta_{a(c)}=\omega_{a(c)}-\omega_{p}/2$, we obtain Eq.~(\ref{eq10}).
            
            Next we apply the (time-dependent) squeezing transformation $\hat{U}_{S}[r(t)]$ (Eq.~(\ref{eq11}))
                \begin{eqnarray}
                    \hat{H}^{S}=\hat{U}_{S}\hat{H}'\hat{U}_{S}^{\dagger}-i\hat{U}_{S}\frac{d}{dt}\hat{U}_{S}^{\dagger},
                    \label{b6}
                \end{eqnarray}
            using
                \begin{eqnarray}
                    \hat{U}_{S}\hat{a}\hat{U}_{S}^{\dagger} &=\hat{a}\cosh r(t)+\hat{a}^{\dagger}\sinh r(t),\nonumber \\
                    \hat{U}_{S}\hat{a}^{\dagger}\hat{U}_{S}^{\dagger} &=\hat{a}\sinh r(t)+\hat{a}^{\dagger}\cosh r(t),
                    \label{b7}
                \end{eqnarray}
            One then finds
                \begin{equation}
                \begin{aligned}
                    \tfrac{\delta_{a}}{2}\hat{\sigma}_z \;&\to\; \tfrac{\delta_{a}}{2}\hat{\sigma}_z,\\[2pt]
                    \delta_{c}\hat{a}^\dagger\hat{a} \;&\to\;
                    \delta_{c}\Bigl[
                        \hat{a}^\dagger\hat{a}\cosh 2r(t)\\
                    &\qquad + \tfrac{\sinh 2r(t)}{2}\bigl(\hat{a}^2+\hat{a}^{\dagger 2}\bigr)\\
                    &\qquad + \sinh^2 r(t)\Bigr],\\[2pt]
                    -\tfrac{\lambda(t)}{2}\bigl(\hat{a}^{\dagger 2}+\hat{a}^2\bigr) \;&\to\;
                    -\lambda(t)\Bigl[
                        \tfrac{\cosh 2r(t)}{2}\bigl(\hat{a}^{\dagger 2}+\hat{a}^2\bigr)\\
                    &\qquad + \sinh 2r(t)\Bigl(\hat{a}^\dagger\hat{a}+\tfrac{1}{2}\Bigr)\Bigr],\\[2pt]
                    g\bigl(\hat{a}^\dagger\hat{\sigma}_{-}+\hat{a}\hat{\sigma}_{+}\bigr) \;&\to\;
                    \tfrac{g}{2}e^{r(t)}(\hat{a}^\dagger+\hat{a})(\hat{\sigma}_{+}+\hat{\sigma}_{-})\\
                    &\quad - \tfrac{g}{2}e^{-r(t)}(\hat{a}^\dagger-\hat{a})(\hat{\sigma}_{+}-\hat{\sigma}_{-}).
                \end{aligned}
                \end{equation}
            The constant energy shift $\delta_{c}\sinh^{2}r(t) - \tfrac{1}{2}\lambda(t)\sinh 2r(t)$ is omitted, as it does not affect the dynamics.
            The resulting transformed Hamiltonian is
                \begin{eqnarray}
                    \hat{H}^{S}=\hat{H}_{\text{Rabi}}+\hat{H}_{\text{err}}+\hat{H}_{\text{DA}},
                    \label{b8}
                \end{eqnarray} 
            where
                \begin{eqnarray}
                    \hat{H}_{\text{DA}} =-\frac{i\dot{r}(t)}{2}(\hat{a}^{\dagger2}-\hat{a}^{2}),
                    \label{b9}
                \end{eqnarray}
            is the drive-amplitude term arising from a time-dependent drive amplitude.

        \section{Semiclassical approximation and Poincaré section}\label{app:C}
            Coherent states minimize quantum uncertainty, providing the closest quantum analog to classical phase-space points.
            Atomic Bloch and photonic Glauber coherent states in Eq.~\ref{eq16} and Eq.~\ref{eq17} correspond to $(q_{1},p_{1})$ and $(q_{2},p_{2})$ in phase-space and satisfy
                \begin{eqnarray}
                    \tau&=&\frac{q_{1}+ip_{1}}{\sqrt{2-q_{1}^{2}-p_{1}^{2}}},\nonumber \\
                    \beta&=&\frac{1}{\sqrt{2}}(q_{2}+ip_{2}),
                    \label{eq18}
                \end{eqnarray}    
            In what follows, we use these coordinates to parameterize the coherent states.
            
            Using the mean-field approximation~\cite{deAguiar1992AP,Wang2025CTP},
                \begin{eqnarray}
                    \bra{\tau}\hat{\sigma}_{+}\ket{\tau}&=&\frac{\tau^{*}}{1+\tau\tau^{*}},\nonumber \\
                    \bra{\tau}\hat{\sigma}_{-}\ket{\tau}&=&\frac{\tau}{1+\tau\tau^{*}},\nonumber \\
                    \bra{\tau}\hat{\sigma}_{z}\ket{\tau}&=&-\frac{1-\tau\tau^{*}}{1+\tau\tau^{*}},\nonumber \\
                    \bra{\beta}\hat{a}\ket{\beta}&=&\beta,\nonumber \\
                    \bra{\beta}\hat{a}^{\dagger}\ket{\beta}&=&\beta^{*},
                    \label{eq19}
                \end{eqnarray}
            The classical effective Hamiltonian becomes
                \begin{eqnarray}
                    \hat{H}_{\text{cl}}&\equiv&\bra{\psi}\hat{H}_{\text{eff}}\ket{\psi}\nonumber \\
                                    &=&\frac{\delta_{a}}{2}(q_{1}^{2}+p_{1}^{2}-1)+\frac{\Omega_{c}}{2}(q_{2}^{2}+p_{2}^{2})\nonumber \\
                                        &&+\tilde{g}q_{1}q_{2}\sqrt{4-2(q_{1}^{2}+p_{1}^{2})}\nonumber \\
                                        &&-\frac{g}{2}e^{-r}p_{1}p_{2}\sqrt{4-2(q_{1}^2+p_{1}^2)},
                    \label{eq20}
                \end{eqnarray}
            where $\ket{\psi}$ is the coherent state defined in Eq.~(\ref{eq15}).

            The classical trajectories in the four-dimensional phase space are obtained by solving Hamilton's equations.
            The Poincaré section is then defined by the condition $q_{2}=0$ with $p_{2}>0$.
            Distribution of points on Poincaré section captures the structure of the full phase-space dynamics.

            We found that the Poincaré sections of both systems are divided into chaotic sea and stable island (see Fig.~\ref{fig1}): the chaotic sea comprises irregularly scattered points, whereas regular trajectories form well-ordered rings at the periphery (points $R_{1(2)}$ and $C_{1(2)}$ denote regular and chaotic trajectories, respectively).
            The distribution on the section reflects the orbits for different initial points in the entire phase-space, increasing $r$ enlarges the chaotic region.
            The sections obtained from the classical effective Hamiltonian $H_{\text{cl}}$ share the same qualitative features as those of the semiclassical QRM.

    \bibliographystyle{apsrev4-2}
    \bibliography{refs}

\end{document}